\documentclass[prb,showpacs,preprintnumbers,aps]{revtex4}
\begin{document}

\title{A Landau fluid model for warm collisionless plasmas}
\author{P. Goswami, T. Passot and P.L. Sulem}
\affiliation{ CNRS, Observatoire de la C\^ote d'Azur, 
B.P. 4229, 06304 Nice Cedex 4, France}

\date{\today}

\begin{abstract}
A Landau fluid model for a collisionless 
electron-proton magnetized plasma, that accurately reproduces the dispersion
relation and the Landau damping rate of all the magnetohydrodynamic waves,
is presented. It is obtained by 
an accurate closure of the hydrodynamic hierarchy at the level of the 
fourth order moments, based on  linear kinetic theory. It retains  non-gyrotropic
corrections to the pressure and heat flux tensors up to the
second order in the ratio between 
the considered frequencies and the ion cyclotron frequency. 

\end{abstract}

\maketitle

\section{Introduction} 

In many spatial and astrophysical plasmas, collisions are negligible, 
making the usual magnetohydrodynamics (MHD) questionable. 
The presence of a strong ambient magnetic field nevertheless ensures a
collective behavior of the plasma, making a hydrodynamic approach of 
the large-scale dynamics possible and even  
advantageous, compared with purely kinetic descriptions 
provided by the Vlasov-Maxwell (VM) or the gyrokinetic equations.
It is thus of great interest, both for the numerical simulation  of broad
spectrum phenomena and for an easier interpretation of the involved processes,
to construct fluid models that extend the MHD equations to
collisionless situations by including finite Larmor radius (FLR) corrections 
and Landau damping. In a fluid formalism, FLR
corrections refer to the part of the pressure and
heat flux tensors associated with the deviation from
gyrotropy. They play a role when the transverse  
scales under consideration extend  up to or beyond the ion Larmor radius 
(fluid models are always limited to parallel scales large
compared with the ion Larmor radius). Evolving on a shorter time scale than the
basic hydrodynamic fields, FLR corrections
can generally be  computed perturbatively. This expansion cannot however be
pushed arbitrary far and any fluid  analysis addressing (transverse) scales
comparable to the ion Larmor radius \cite{PSH98} can only be heuristic.

 From Vlasov equation it is easy to derive a set of exact moment equations.
This fluid hierarchy is however faced with a closure
problem. An interesting approach consists in closing this
hierarchy  by using relations, derived from linearized kinetic theory,
between  given moments and lower order 
ones. This in particular accounts for linear Landau damping  in a fluid
formalism. Such an approach initiated in Ref. [2] leads to descriptions
usually referred to as Landau fluids. We here concentrate on a closure at the
level of the fourth order moments, which provides an accurate description of
most of the usual hydrodynamic quantities.

An alternative method to the Landau fluids is provided by the
gyrofluids \cite{HL92,SH01}
obtained by taking the moments of gyrokinetic equations. The same closure
problem is encountered for the moment hierarchy. The gyrofluids have the advantage of 
retaining FLR corrections to all order relatively to the transverse scale
within a low frequency asymptotics but, being written in a local reference
frame, the resulting equations  are more complex than those governing 
the Landau fluids, we are here concerned with. 

As an example, Landau fluid models should be most useful to analyze the
dynamics of the magnetosheath that appears as a buffer  between the earth bow 
shock and the magnetopause and  plays an important role in decreasing
the impact of solar activity on the earth environment. Recent analyses of
data provided by the Cluster spacecraft mission have revealed that the 
magnetosheath displays a wide spectrum of low frequency modes (Alfv\'en, slow
and fast magnetosonic, mirror) \cite{SPB03} whose wavelengths extend down to the
ion gyroradius and beyond. Since the plasma is relatively warm and
collisionless,  Landau damping and FLR corrections are
supposed to play an important role. 
Coherent solitonic structures (magnetic holes and shocklets) are also
observed, and their origin is still debated. \cite{S04b,T05}

A Landau fluid model for collisionless purely magnetohydrodynamic regimes 
\cite{SHD97} was first derived  from the equation for the distribution
function of the particle guiding centers, taken to lowest order. It is thus 
restricted to the largest MHD scales where the pressure and heat flux tensors for 
each species  can be viewed as gyrotropic and where the transverse velocity reduces 
to the electric drift. Starting directly from the VM
equations, this model was then extended in order to include a generalized Ohm's
law and to retain the leading order FLR corrections to the pressure tensor.
\cite{PS03b, BPS04} This model enabled one to reproduce the
dynamics of dispersive Alfv\'en 
waves propagating along the ambient field both in the linear and
weakly-nonlinear regimes and to recover the  
kinetic derivative nonlinear Schr\"odinger (KDNLS)
equation in a long-wave asymptotic expansion with, as the only difference, 
the replacement of the plasma response function by its two or
four poles Pad\'e approximants. It also accurately describes the
dissipation of oblique magnetosonic waves. \cite{BGPS05} Non-gyrotropic
contributions to the heat fluxes were introduced in Ref. [12] in order to
obtain the dispersion relation and the Landau damping rate 
of oblique and kinetic Alfv\'en waves.
The approach we present here provides  a more systematic description of the FLR
corrections up to second order, by retaining parallel and transverse
heat flux vectors whose coupling to the non-gyrotropic pressure
contributions is in particular required for an accurate description
of the transverse magnetosonic waves. \cite{MS85}
A recent paper by Ramos \cite{R05} addresses a similar issue and
derives a complete set of nonlinear equations for fluid moments up
to the heat flux vectors, leaving the closure on the
fourth order moments unspecified. We here follow a similar path choosing in
Section II to linearize the equations for the (``slaved'') non-gyrotropic
contributions to the pressure and heat flux tensors, while retaining nonlinear
equations for all the other moments.  While Ramos performs a first order
expansion in the regime referred to as the fast dynamics ordering,  we here keep the second
order accuracy necessary for a proper description of the
oblique dynamics. By fitting with the kinetic
theory briefly reported in Section III,  we also give in Section IV an
explicit closure relation, taking into account FLR corrections, and
approximating the plasma response function with four and three poles
Pad\'e approximants in order to recover accurate limits for Landau
damping both in the isothermal and adiabatic regimes. As the result of such a
high order approximation, 
one of the fourth order moments is prescribed as the solution of a
dynamical equation. 
After a  discussion of the resulting model in Section V, the
validation of the model at the level of the dispersion relation of the
various MHD waves is addressed in Section VI. Section VII  is the conclusion
where further extensions to a model, aimed at including a realistic description
of the mirror modes, are announced.

\section{Fluid description of each particle species} 

\subsection{The moment hierarchy}

Starting from  the VM equations for 
the distribution function $f_r$  of the particles of species $r$ with charge $q_r$,
mass $m_r$, and  average number density $n_r$, one easily derives a
hierarchy of fluid equations 
for the corresponding density $\rho_r = m_r n_r\int f_r d^3v$, hydrodynamic velocity 
${ u_r ={\int v f_r d^3v}/{\int f_r d^3v}}$, 
pressure tensor    
${ {\bf p}_r = 
m_r n_r \int (v-u_r)\otimes (v-u_r)  f_r d^3 v}$ and  heat flux tensor
 ${ {\bf q}_r=m_r n_r \int (v-u_r)\otimes (v-u_r) \otimes (v-u_r) f_r d^3 v} $,
 in the usual form 
\begin{eqnarray}
&&\partial_t \rho_r + \nabla \cdot (\rho_r u_r) =0 \label{densityr}\\ 
&&\partial_t u_r + u_r \cdot \nabla u_r + \frac{1}{\rho_r} \nabla \cdot 
{\bf p}_r - \frac{q_r}{m_r}(E + \frac{1}{c} u_r \times B) = 0 \label{ur}\\ 
&&\partial_t {\bf p}_r + \nabla \cdot (u_r {\bf p}_r +  {\bf q}_r) +
 \Big [{\bf p}_r  \cdot \nabla  u_r +
\frac{q_r}{m_r c} B \times {\bf p}_r \Big ]^{\cal S}=0, \label{pressionr}
\end{eqnarray}
where the tensor ${\displaystyle B \times {\bf p}_r}$ has elements 
${\displaystyle (B \times {\bf p}_r)_{ij} = \epsilon_{iml}B_m p_{r \,lj}}$ 
and where, for a square matrix ${\displaystyle{\bf a}}$, one defines
${\displaystyle {\bf a}^{\cal S} = {\bf a} + {\bf a}^{\rm tr}} $. 
One has  ${\displaystyle(B \times {\bf p}_r)^{\rm tr}}=-{\bf p}_r\times B$.
In order to distinguish between scalar and tensorial pressures, bold letters are
used to denote tensors of rank two and higher. The equation for the heat flux
tensor involves the fourth order moment 
${{\bf r}_r=m_r n_r \int (v-u_r)\otimes (v-u_r) \otimes (v-u_r) 
\otimes (v-u_r) f_r d^3 v} $.
Since at this step we are dealing with the
various particle species separately, we simplify the
writing by hereafter dropping the $r$ subscript. 
The equations governing the heat flux  elements then read
\begin{eqnarray}
&&\partial_t q_{ijk} + v_l \partial_l q_{ijk} + \partial_l r_{ijkl} -
\frac{1}{\rho} \partial_l p_{lm}(\delta_{mi} p_{jk}+ \delta_{mj} p_{ik} +
\delta_{mk} p_{ij})\nonumber \\
&&+ \partial_l u_m (\delta_{mi} q_{jkl} + \delta_{mj}q_{ikl} + \delta_{mk}q_{ijl}
  + \delta_{ml}q_{ijk}) - \Omega {\widehat b_n} (\epsilon_{imn} q_{jkm}  + \epsilon_{jmn} q_{ikm}  + 
\epsilon_{kmn} q_{ijm}) = 0. \nonumber \\
\label{heatflux}
\end{eqnarray}
We here concentrate on the ion dynamics. The corresponding
equations for the electrons are obtained from the equations for the ions by changing the sign of the
electric charge (including in the cyclotron frequency) and making
the approximation $m_e/m_p \ll 1$.

\subsection{Pressure tensors and heat flux vectors}

In order to isolate the gyrotropic components of the pressure tensor,
it is convenient to rewrite Eq. (\ref{pressionr}) for 
the pressure tensor of each particle species in the form 
\begin{equation} 
{\bf p}\times {\widehat b} - {\widehat b} \times {\bf p} = {\bf k}
\label{Ptensor2} 
\end{equation}
where $ \displaystyle{{\widehat b} = \frac{B}{|B|}}$ is the unit vector along 
the local magnetic field and
\begin{equation} 
{\bf k}= \frac{1}{\Omega} \frac{B_0}{|B|} \Big [
\frac{d {\bf p}}{dt} + (\nabla \cdot u){\bf p} + \nabla\cdot {\bf q} +
({\bf p}  \cdot \nabla  u)^{\cal S}\Big ].
\end{equation}
In this equation, $B_0$ denotes the amplitude of the
ambient field assumed to be oriented in the $z$-direction,
  and  $\displaystyle{\Omega= \frac
{q B_0}{m c}} $ is the cyclotron frequency of the considered particles species 
with charge $q$  and mass $m$. Furthermore, 
${\displaystyle \frac{d}{dt} = \partial_t + u \cdot \nabla}$ denotes the 
convective derivative.

We first note that the left-hand side of Eq. (\ref{Ptensor2}) can be viewed
as a self-adjoint linear operator 
acting on ${\bf p}$, whose kernel is spanned by the tensors
${\displaystyle {\bf n} = {\bf I} -
{\widehat b}\otimes {\widehat b}}$ and  ${\displaystyle  \mbox{\boldmath
$\tau$}={\widehat b}\otimes {\widehat b}}$. Using the symbol $:$ to denote
double contraction, it is convenient to 
define the projection $\displaystyle{
{\overline{\bf a}}}$ of any $(3\times 3)$ rank two tensor
 ${\bf a}$  on the image of this operator as
$\displaystyle{
{\overline{\bf a}} =   {\bf a} - \frac{1}{2} ({\bf a} : {\bf n}) {\bf n}  - 
({\bf a} : \mbox{\boldmath$\tau$})  \mbox{\boldmath $\tau$}}$,
which implies  $\displaystyle{{\rm tr} \, {\overline{\bf a}} = 0 }$  and
$\displaystyle{{\overline{\bf a}}:\mbox{\boldmath $\tau$} = 0}$. 
In particular, the pressure tensor ${\displaystyle {\bf p} =  {\bf P} +
\mbox{\boldmath$\Pi$} }$  is written as the sum of a 
gyrotropic pressure  
${\displaystyle {\bf P} =  p_{\perp }  {\bf n} + p_{\| }
  \mbox{\boldmath$\tau$} } $ (with  
${\displaystyle 2 p_\perp = {\bf p} : {\bf n}}$ and
${\displaystyle p_\|  ={\bf p} :\mbox{\boldmath $\tau$}}$) and of 
a gyroviscous stress  ${\displaystyle \mbox{\boldmath $\Pi$}
= {\overline{{\bf p}}}}$ that satisfies 
${\displaystyle  \mbox{\boldmath $\Pi$}: {\bf n}= 0}$  and 
${\displaystyle 
 \mbox{\boldmath $\Pi$}: \mbox{\boldmath $\tau$} = 0}$.

%\subsubsection{The heat flux vectors}

A similar decomposition is performed on the heat flux tensor by writing ${\bf q} =
{\bf S} +  \mbox{\boldmath $\sigma$}$ with the conditions
$\sigma_{ijk} n_{jk} = 0$ and $\sigma_{ijk} \tau_{jk} = 0$. One has  
\begin{eqnarray}
&& S_{ijk} = \frac{1}{2} \Big ( S_i^\perp n_{jk} +  S_j^\perp n_{ik} +  S_k^\perp
n_{ij} +  S_l^\perp \tau_{li}n_{jk} +  S_l^\perp \tau_{lj}n_{ik} +
S_l^\perp \tau_{lk}n_{ij} \Big ) \nonumber \\
&&+ S_i^\|\tau_{jk} + S_j^\|\tau_{ik} + S_k^\|\tau_{ij} - \frac{2}{3} \Big
( S_l^\|\tau_{li} \tau_{jk} + S_l^\|\tau_{lj} \tau_{ik} + S_l^\|\tau_{lk}
\tau_{ij} \Big ),
\end{eqnarray}
where the parallel and transverse heat flux vectors $S^\|$ and  $S^\perp$ have components
$S_i^\| = q_{ijk} \tau_{jk}$ and $2 S_i^\perp = q_{ijk} n_{jk}$. In the
special case where the tensor ${\bf q}$ is gyrotropic, only the
$z$-components $q_\|= S_\| \cdot {\widehat b}$ and 
$q_\perp= S_\perp \cdot{\widehat b} $ are non zero. Transverse components are
however required,  for example to describe transverse magnetosonic waves. 
\cite{MS85}

We consider in this paper perturbations that are at large scale in all space
directions and in time, with an amplitude that is relatively small. This leads
us to retain the terms involving the non-gyrotropic parts of the pressure and
heat flux tensors at the linear level only. Such an ordering implies in
particular that increasing the amplitude of the fluctuations requires longer
length scales for preserving a given accuracy.
In the following, we shall thus neglect the $\sigma$ contribution to the heat
flux tensor. One indeed easily checks from the equation satisfied by $\mbox{\boldmath
  $\sigma$}$ (see Appendix 2 of Ref. [15]) that   $\mbox{\boldmath $\sigma$}$ 
involves either  nonlinear contributions or linear contributions of
second order relatively to the scale separation parameter, and  thus
turns out to be negligible in the equations for the gyroviscous stress
or for the heat fluxes, at the order of the present analysis. 

\subsection{Dynamics of the gyrotropic pressures}

To obtain the equations for the gyrotropic pressure components, one applies
the contraction with the tensors $\bf{I}$ and
$\mbox{\boldmath $\tau$}$ on both sides of
Eq. (\ref{Ptensor2}) to get \cite{OCF68,PS04}
\begin{eqnarray}
&&\partial_t p_{\perp } + \nabla\cdot(u \, p_{\perp })+ 
p_{\perp } \nabla\cdot u - p_{\perp }\,{\widehat b}\cdot\nabla u \cdot
{\widehat b}  + \frac{1}{2}\Big ({\rm tr}\,\nabla \cdot{\bf q}  - 
{\widehat b}\cdot (\nabla \cdot{\bf q}) \cdot  {\widehat b} \Big )
\nonumber \\
&&\qquad \ +\frac{1}{2}\Big (
{\rm tr}\, (\mbox{\boldmath $\Pi$}\cdot \nabla u)^S -
(\mbox{\boldmath $\Pi$} \cdot \nabla u)^S :  \mbox{\boldmath $\tau$} +
\mbox{\boldmath $\Pi$} :\frac{d \mbox{\boldmath $\tau$} }{dt}
\Big )=0 \label{p_perp}\\
&&\partial_t p_{\|} + \nabla \cdot (u \, p_{\|}) + 2  p_{\|}\,
{\widehat b}\cdot \nabla u\, \cdot{\widehat b} + 
{\widehat b}\cdot (\nabla \cdot{\bf q}) \cdot {\widehat b} +
(\mbox{\boldmath $\Pi$} \cdot \nabla u)^S : \mbox{\boldmath $\tau$}  -
\mbox{\boldmath $\Pi$} :\frac{d \mbox{\boldmath $\tau$} }{dt}
=0, \label{p_paral}
\end{eqnarray}
which appear as the condition for the solvability of Eq. (\ref{Ptensor2}).
Note that it is important to retain the coupling to the
gyroviscous stress (in spite of its smallness) in order to ensure energy conservation whatever the
form of the forthcoming closure relations.\cite{R05}

Since  $ \mbox{\boldmath $\sigma$}$
does not contribute at a linear level in the pressure equations, we can 
neglect it and write
\begin{eqnarray}
&&{\widehat b}\cdot (\nabla \cdot{\bf q}) \cdot  {\widehat b} \approx  
 -2 ({\widehat b} \cdot S^\perp)\nabla \cdot {\widehat b} + \nabla  \cdot S^\|
-2{\widehat b} \cdot \nabla  {\widehat b }\cdot S^\| \\
&& \frac{1}{2} \Big ( {\rm tr} ( \nabla \cdot{\bf q} ) - 
{\widehat b}\cdot (\nabla \cdot{\bf q}) \cdot  {\widehat b} \Big )\approx
 \nabla \cdot S^\perp + ({\widehat b} \cdot S^\perp)\nabla \cdot {\widehat
b} + {\widehat b} \cdot \nabla  {\widehat b }\cdot S^\|.
\end{eqnarray}

\subsection{Gyroviscous stress tensor}

In order to determine the non-gyrotropic contributions to the pressure tensor
of the various particle species, we start from Eq. (\ref{Ptensor2}) for the
full pressure tensor.
Using Eqs. (\ref{p_perp})-(\ref{p_paral}) governing the gyrotropic 
pressures,  Eq. (\ref{Ptensor2})  is rewritten 
\begin{equation} 
 \mbox{\boldmath $\Pi$} \times {\widehat b} - 
{\widehat b} \times  \mbox{\boldmath $\Pi$} = {\overline{\mbox{\boldmath
      $\kappa$}}} + {\overline{L(\mbox{\boldmath $\Pi$})}}   \label{pi}
\end{equation}
where
\begin{equation}
\mbox{\boldmath $\kappa$} =  \frac{1}{\Omega} \frac{B_0}{|B|} \Big [
\frac{d {\bf P}}{dt} + (\nabla \cdot u){\bf P } + 
\nabla\cdot {\bf q} +
({\bf P}  \cdot \nabla  u)^{\cal S}\Big ] \label{kappar}
\end{equation}
and 
\begin{equation}
L(\mbox{\boldmath $\Pi$}) = \frac{1}{\Omega}  \frac{B_0}{|B|}
\Big [
\frac{d \mbox{\boldmath $\Pi$}}{dt} + 
(\nabla \cdot u)\mbox{\boldmath $\Pi$} + 
(\mbox{\boldmath $\Pi$}  \cdot \nabla  u)^{\cal S}\Big ]. 
\label{L(pi)}
\end{equation}
%In $\overline{\mbox{\boldmath $\kappa$}}$, the second term of the r.h.s. of Eq.
%(\ref{kappar}) does not contribute, while the first one rewrites 
%${\displaystyle
%\overline{\frac{d {\bf p}^G}{dt}}=(p_{\| }- p_{\perp })\frac{d\mbox{\boldmath $\tau$}}{dt}}$.
The elements of $\mbox{\boldmath $\kappa$}$ rewrite
\begin{eqnarray}
&&\overline{\kappa}_{ij} =  \frac{1}{\Omega} \frac{B_0}{|b|} \Big [(p_\| - p_\perp)
\frac{d\tau_{ij}}{dt} +  \overline{ \partial_k q_{kij}} + p_\perp
(  n_{ik} \partial_k u_j  + n_{jk}\partial_k u_i -  n_{ij}n_{kl}\partial_l u_k)
 \nonumber \\
&&+ p_\|(  \tau_{ik} \partial_k u_j  + \tau_{jk}\partial_k u_i - 
2 \tau_{ij} \tau_{kl}\partial_l u_k )\Big ].
\end{eqnarray}
Furthermore in Eq. (\ref{pi}), the element of the left-hand side
with ${ij}$ indices reads $\displaystyle{
\epsilon_{jkl} \Pi_{ik} b_l - \epsilon_{ikl} b_k \Pi_{lj} =
b_l(\epsilon_{jkl}\Pi_{ik} +  \epsilon_{ikl}\Pi_{kj})}$, thus suggesting a
misprint in Eq. (3.5) of Ref. [15].
When neglecting as previously the contribution originating from
$\mbox{\boldmath $\sigma$}$,  the heat flux term 
$\displaystyle{ {\overline{ \partial_k q_{kij}}}}$ reduces to 
\begin{equation}
 \overline{ \partial_k q_{kij}} \approx \Big (\overline {\nabla \cdot S }\Big )_{ij} =
\partial_k S_{kij} - \frac{1}{2} n_{ij} n_{mn}  \partial_k S_{kmn} - 
  \tau_{ij} \tau_{mn} \partial_k S_{kmn}. 
\end{equation}

In the linear approximation, we have
\begin{eqnarray}
&&\overline{(\nabla\cdot {\bf S})}_{ij} = \frac{1}{2} \Big [\partial_k \Big (S^\perp_i +
(S^\perp\cdot{\widehat b}) \,{\widehat b_i}\Big )n_{jk} + 
\partial_k \Big (S^\perp_j +
(S^\perp\cdot{\widehat b}) \, {\widehat b_j}\Big )n_{ik}\Big ] \nonumber \\
&& + {\widehat b_j}  ({\widehat b}\cdot \nabla) S^\|_i + 
{\widehat b_i}  ({\widehat b}\cdot \nabla) S^\|_j 
- 2 ({\widehat b}\cdot  \nabla S^\|\cdot {\widehat b}) \tau_{ij}
- \frac{1}{2} (\nabla \cdot S^\perp - 
{\widehat b}\cdot  \nabla S^\perp\cdot {\widehat b}) n_{ij}
\end{eqnarray}
where the derivatives act only on the heat flux components. This yields
(the superscript $(0)$ refers to equilibrium quantities)
\begin{eqnarray}
&&\partial_t \Pi_{xx} - 2 \Omega \Pi_{xy} + p_\perp^{(0)} (\partial_x u_x -
\partial_y u_y) + \frac{1}{2} (\partial_x S_x^\perp- \partial_y S_y^\perp) =0 \label{pixy}\\
&&\partial_t \Pi_{xy} + 2 \Omega \Pi_{xx} + p_\perp^{(0)} (\partial_x u_y +
\partial_y u_x) + \frac{1}{2} (\partial_y S_x^\perp+ \partial_x S_y^\perp) =\label{pixx}0\\
&&\partial_t \Pi_{xz} - \Omega \Pi_{yz} + p_\perp^{(0)} \partial_x u_z +
p_\|^{(0)}\partial_z u_x + \partial_x S_z^\perp+ \partial_z S_x^\| 
- (p_\perp^{(0)} -  p_\|^{(0)}) \partial_t {\widehat b_x}=0\label{piyz}\\
&&\partial_t \Pi_{yz} + \Omega \Pi_{xz} + p_\perp^{(0)} \partial_y u_z +
p_\|^{(0)}\partial_z u_y + \partial_y S_z^\perp+ \partial_z S_y^\| 
- (p_\perp^{(0)} -  p_\|^{(0)}) \partial_t {\widehat b_y}=0\label{pixz}
\end{eqnarray}
together with $\displaystyle{\Pi_{xx} = - \Pi_{yy} }$ and
$\displaystyle{\Pi_{zz} =0}$.
Defining the transverse divergence of the
gyroviscous stress  tensor $\displaystyle{\nabla_\perp \cdot \Pi_\perp}$
as the vector of components $\displaystyle{\Big (\partial_x \Pi_{xx} +\partial_y
  \Pi_{xy}, \partial_x \Pi_{xy} +\partial_y
  \Pi_{yy} , 0\Big )}$ and introducing the unit vector $\displaystyle{{\widehat z}}$ in the
direction of the ambient field, Eqs. (\ref{pixy}) and (\ref{pixx}) then give
\begin{equation}
   \nabla_\perp \cdot \Pi_\perp + \frac{1}{4\Omega} \Delta_\perp S^\perp\times {\widehat z}
  = - \frac{p_\perp^{(0)}}{2\Omega} \Delta_\perp u \times {\widehat z} -
\frac{1}{2\Omega} \partial_t \left (\nabla_\perp \cdot \Pi_\perp \right
  )\times  {\widehat z}. \label{divpiperp}
\end{equation}
On the other hand, defining the vector $\displaystyle{\Pi_z = (\Pi_{xz},
  \Pi_{yz}, \Pi_{zz}=0 )}$, Eqs.  (\ref{piyz}) and (\ref{pixz}) rewrite
\begin{equation}
-\Omega \Pi_z \times {\widehat z} + \partial_z S^\|_\perp = 
- \nabla_\perp S^\perp_z - p_\perp^{(0)} \nabla_\perp u_z - p_\|^{(0)}
\partial_z u_\perp + \Big ( p_\perp^{(0)} - p_\|^{(0)} \Big ) \partial_t
{\widehat b_\perp} - \partial_t \Pi_z. \label{piz}
\end{equation}

\subsection{Dynamics of the heat flux vectors}

Equation (\ref{heatflux}) for the heat flux tensor involves the divergence of
the fourth order moment ${\bf r}$, that at this step should be simplified in
order to conveniently close the hierarchy at the present order.
We first note that 
instead of dealing with the fourth order moment ${\bf r}$, it is convenient to
isolate the deviation from the product of second order moments by writing  
\begin{eqnarray}
&&\rho r_{ijkl} =  P_{ij}P_{lk} + P_{ik}P_{jl} + P_{il}P_{jk} +  
P_{ij}\Pi_{lk} + P_{ik}\Pi_{jl} + P_{il}\Pi_{jk} \nonumber\\ 
&& +  \Pi_{ij}P_{lk} + \Pi_{ik}P_{jl} + \Pi_{il}P_{jk} + \rho {\widetilde r_{ijkl}}.
\end{eqnarray} 
The correction term $\displaystyle{\rho {\widetilde r_{ijkl}}}$ a priori
includes a contribution of the form 
$\displaystyle{ \Pi_{ij}\Pi_{lk} + \Pi_{ik}\Pi_{jl} + \Pi_{il}\Pi_{jk}}$ 
that we here neglect since, as already mentioned, contributions from the gyroviscous
stress are retained in linear terms only (except in Eqs. (\ref{p_perp}) and 
 (\ref{p_paral}) in order to ensure energy conservation). This 
 algebraic transformation allows significant simplifications in the
 forthcoming equations. Second, we
 make the approximation of retaining only the gyrotropic part of the tensor
 ${\widetilde {\bf r}}$ that is then given by 
\begin{eqnarray}
&&{\widetilde r_{ijkl}} = \frac{{\widetilde r_{\|\|}}}{3} (\tau_{ij}\tau_{kl}+
\tau_{ik}\tau_{jl} + \tau_{il}\tau_{jk} )
+ {\widetilde r_{\|\perp}}(n_{ij}\tau_{kl}+ n_{ik}\tau_{jl} + n_{il}\tau_{jk} \\
&&+ \tau_{ij}n_{kl}+ \tau_{ik}n_{jl} + \tau_{il}n_{jk} )
 + \frac{{\widetilde r_{\perp\perp}}}{2}(n_{ij}n_{kl}+ n_{ik}n_{jl} + n_{il}n_{jk}).
\end{eqnarray}
The scalar quantities
$\displaystyle {r_{\|\|}= r_{ijlk} \tau_{ij}\tau_{kl}}$, 
 $\displaystyle{r_{\perp\|}= \frac{1}{2} r_{ijlk}n_{ij}\tau_{kl}}$
and  $\displaystyle{r_{\perp\perp}=  \frac{1}{4}r_{ijlk} n_{ij}n_{kl}}$ 
are related to  $\displaystyle{{\widetilde r_{\|\|}}}$,  $\displaystyle{{\widetilde
    r_{\|\perp}}}$ and 
$\displaystyle{{\widetilde r_{\perp\perp}}}$  (given by similar formulas with $\displaystyle{r_{ijkl}}$
replaced by  $\displaystyle{{\widetilde{r}_{ijkl}}}$) by 
\begin{eqnarray}
&& {\widetilde r_{\|\|}} =r_{\|\|} - 3 \frac{p_\|^2}{\rho} \label{rparpar} \\
&& {\widetilde r_{\|\perp}} =r_{\|\perp} -  \frac{p_\perp p_\|}{\rho} \label{rparperp}\\
&& {\widetilde r_{\perp\perp}} =r_{\perp\perp} -
  2\frac{p_\perp^2}{\rho} \label{rperpperp} .
\end{eqnarray}

One derives the equations for the
heat flux vectors by writing $\displaystyle{\frac{dS^\|_i}{dt} =
S_{ijk} \frac{d\tau_{jk}}{dt} +  \frac{dS_{ijk}}{dt}\tau_{jk} }$ and 
$\displaystyle{\frac{dS^\perp_i}{dt} =
-S_{ijk} \frac{d\tau_{jk}}{dt} +  \frac{dS_{ijk}}{dt}n_{jk} }$.
The first term in the above equations is given by
\begin{equation}
S_{ijk} \frac{d\tau_{jk}}{dt} =  2 (S^\perp - S^\|) \cdot {\widehat b} 
\frac{d{\widehat b_i}}{dt} + 2 S_j^\| \frac{d\tau_{ij}}{dt}
\end{equation}
and the second terms are computed using the dynamical equation for the third
order moment.
One gets
\begin{eqnarray}
&&\frac{dS^\perp_i}{dt}= - (S^\perp - S^\|)\cdot {\widehat b} \frac{d{\widehat
    b_i}}{dt} -  S_j^\| \frac{d\tau_{ij}}{dt} -
\frac{1}{2\rho} (P_{ij}P_{kl} + P_{ik}P_{jl} + P_{il}P_{jk}\nonumber \\
&& + P_{ij}\Pi_{kl} + P_{ik}\Pi_{jl} + P_{il}\Pi_{jk} +
\Pi_{ij}P_{kl} + \Pi_{ik}P_{jl} + \Pi_{il}P_{jk}) \partial_l \tau_{jk}\nonumber \\
&& - (P_{jl} + \Pi_{jl} )\partial_l \Big (\frac{p^\perp}{\rho} (\tau_{ij} + 2
    n_{ij})\Big )   - P_{jl}\partial_l\Big
    (\frac{1}{\rho}n_{jk}\Pi_{ik}\Big ) + \frac{1}{\rho} \Pi_{ik}n_{jk}
    \partial_l\Pi_{jl}  \nonumber \\  
&& - (S^\perp \cdot \nabla) u_i - (\nabla \cdot u) S_i^\perp - \frac{1}{2}
\partial_l u_j \Big ( S_i^\perp n_{jl} +  S_m^\perp n_{mj}n_{il} + S_l^\perp
    n_{ij}  \nonumber \\  
&& +  S_m^\perp \tau_{mi} n_{jl} +  S_m^\perp \tau_{ml}n_{ij} + 
2 S_k^\| \tau_{il}  n_{jk} \Big ) + 
\Omega \epsilon_{ijl}S_j^\perp{\widehat b_l} - \frac{1}{2} n_{jk} \partial_l
   {\widetilde r_{ijkl}} \label{S_perp}
\end{eqnarray}
and
\begin{eqnarray}
&&\frac{dS^\|_i}{dt}= 2 (S^\perp - S^\|)\cdot {\widehat b} \frac{d{\widehat
    b_i}}{dt} + 2 S_j^\| \frac{d\tau_{ij}}{dt} + 
\frac{1}{\rho} (P_{ij}P_{kl} + P_{ik}P_{jl} + P_{il}P_{jk}\nonumber \\
&& + P_{ij}\Pi_{kl} + P_{ik}\Pi_{jl} + P_{il}\Pi_{jk} +
\Pi_{ij}P_{kl} + \Pi_{ik}P_{jl} + \Pi_{il}P_{jk}) \partial_l
    \tau_{jk}\nonumber \\
&& - (P_{jl} + \Pi_{jl} )\partial_l \Big (\frac{p_\|}{\rho} (n_{ij} + 3
    \tau_{ij})\Big ) - 2P_{jl}\partial_l\Big
    (\frac{1}{\rho}\tau_{jk}\Pi_{ik}\Big ) + \frac{2}{\rho} \Pi_{ik}\tau_{jk}
    \partial_l\Pi_{jl} \nonumber \\
&& - (S^\| \cdot \nabla) u_i - (\nabla \cdot u) S_i^\| - 2 \partial_l u_j \Big
    (S^\perp_k \tau_{jk}n_{il} + S^\|_i \tau_{jl} + S^\|_l \tau_{ij} - 
(S^\|\cdot {\widehat b})\tau_{il}{\widehat b}_j\Big )\nonumber \\
&& + \Omega \epsilon_{ijl}S_j^\| {\widehat b_l} - \tau_{jk} \partial_l 
    {\widetilde r_{ijkl}} \label{S_paral}
\end{eqnarray}
which do not totally identify with the result of Ref. [15].

\subsection{Second order approximation of the non-gyrotropic pressures and heat fluxes}

Noting by inspection of Eqs. (\ref{S_perp}) and (\ref{S_paral}) that the
magnitude of the transverse components of the heat flux
vectors  scales proportionally to the inverse gyrofrequency of the ions, we linearize the equations 
for these quantities, while we retain the nonlinear dynamics of the
longitudinal components (see Section II G). Using 
$\displaystyle{\partial_l({\widetilde r_{ixxl}}+ {\widetilde r_{iyyl}})=
2\partial_i {\widetilde r_{\perp\perp}}}$ for $i=x$ or $y$ and
 $\displaystyle{\partial_l{\widetilde r_{izzl}}
= \partial_i{\widetilde r_{\|\perp}}}$, 
and introducing the temperatures $T_\|= m p_\|/\rho$
and $T_\perp= m p_\perp/\rho$ where $m$ is the mass of 
the considered particles, one has 
\begin{equation}
\frac{T_\perp^{(0)}}{m} \nabla_\perp \cdot \Pi_\perp - \Omega S_\perp^\perp \times
{\widehat z} = -2 \frac{p_\perp^{(0)}}{m} \nabla_\perp T_\perp^{(1)} -
2\nabla_\perp {\widetilde r_{\perp\perp}}  -\partial_t S^\perp_\perp \label{Sperp}.
\end{equation}
Similarly,
\begin{equation}
2\frac{T_\|^{(0)}}{m}\partial_z \Pi_{z} - \Omega S_\perp^\| \times {\widehat z} = 
-  \frac{p_\perp^{(0)}}{m}\nabla_\perp T_\|^{(1)} - 
2 \frac{p_\|^{(0)} -  p_\perp^{(0)}}{m} T_\|^{(0)} \partial_z {\widehat
  b_\perp} - \nabla_\perp{\widetilde r_{\|\perp}}  - \partial_t  S_\perp^\|. \label{Sparal}
\end{equation}

Combining Eqs. (\ref{divpiperp}) and (\ref{Sperp}) and defining the square
Larmor radius 
$\displaystyle{r_L^2 = \frac{T_\perp^{(0)}}{m\Omega^2}}$ gives 
\begin{eqnarray}
&&\Big (1 + \frac{1}{4} r_L^2 \Delta_\perp \Big )\nabla_\perp \cdot \Pi_\perp
=   \frac{p_\perp^{(0)}}{2\Omega}{\widehat z}\times
\Delta_\perp u   - \frac{\rho^{(0)}}{2m}r_L^2
\Delta_\perp \nabla_\perp T_\perp^{(1)} \nonumber\\
&& \qquad \qquad - \frac{1}{2\Omega^2}
\Delta_\perp\nabla_\perp {\widetilde r_{\perp\perp}} - 
\frac{1}{2\Omega} \partial_t\Big (\nabla_\perp \cdot
\Pi_\perp \times {\widehat z} + \frac{1}{2\Omega}\Delta_\perp
S^\perp_\perp\Big ) \label{A} \\
&&\Big (1 + \frac{1}{4} r_L^2 \Delta_\perp \Big ) S^\perp_\perp = 
\Big ( \frac{2 p_\perp^{(0)}}{m\Omega} {\widehat z} \times\nabla_\perp T_\perp^{(1)} -
\frac{p_\perp^{(0)}}{2} r_L^2 \Delta_\perp u_\perp + \frac{2}{\Omega} {\widehat z} 
\times\nabla _\perp {\widetilde r_{\perp\perp}} \Big ) \nonumber \\
&& \qquad\qquad  - \partial_t \Big (\frac{ r_L^2}{2}  \nabla_\perp \cdot
\Pi_\perp -  \frac{1}{\Omega} {\widehat z}\times S^\perp \Big ). \label{B}
\end{eqnarray}
Similarly, combining Eqs. (\ref{piz}) and (\ref{Sparal}) gives 
\begin{eqnarray}
&&\Big ( 1 + 2 \frac{T_\|^{(0)}}{m\Omega^2}\partial_{zz} \Big )\Pi_z = 
\frac{{\widehat z}}{\Omega}\times \Big (\nabla_\perp S^\perp_z +
p_\perp^{(0)} \nabla_\perp u_z + p_\|^{(0)}
\partial_z u_\perp - ( p_\perp^{(0)} - p_\|^{(0)} ) \partial_t
{\widehat b_\perp} + \partial_t \Pi_z \Big ) \nonumber \\
&& \qquad - \frac{1}{\Omega^2} \partial_z \Big ( \frac{p_\perp^{(0)}}{m}\nabla_\perp T_\|^{(1)} - 
2 \frac{p_\perp^{(0)} -  p_\|^{(0)}}{m} T_\|^{(0)} \partial_z {\widehat
  b_\perp} + \nabla_\perp{\widetilde r_{\|\perp}} +  \partial_t
S_\perp^\|\Big ) \label{C}\\
&&\Big ( 1 + 2 \frac{T_\|^{(0)}}{m\Omega^2}\partial_{zz} \Big )S_\perp^\| =
\frac{\widehat z}{\Omega} \times \Big (\frac{p_\perp^{(0)}}{m}\nabla_\perp T_\|^{(1)} -
2 \frac{p_\perp^{(0)} -  p_\|^{(0)}}{m} T_\|^{(0)} \partial_z {\widehat
  b_\perp} + \nabla_\perp{\widetilde r_{\|\perp}}  +\partial_t  S_\perp^\|
\Big ) \nonumber \\
&& \qquad - \frac{2 T_\|^{(0)}}{m\Omega^2} \partial_z 
\Big ( \nabla_\perp S^\perp_z + p_\perp^{(0)} \nabla_\perp u_z + p_\|^{(0)}
\partial_z u_\perp - \Big ( p_\perp^{(0)} - p_\|^{(0)} \Big ) \partial_t
{\widehat b_\perp} - \partial_t \Pi_z\Big ). \label{D}
\end{eqnarray}
Note that the operators in the l.h.s. of eqs. (\ref{A})-(\ref{D}) cannot be
inverted for any wavenumber, indicating the limitation of the fluid approach
to large scales, both in the longitudinal and transverse directions.
At second order in terms of 
$\displaystyle{  \frac{\omega}{\Omega} 
\sim  \sqrt{\frac{2T_\|^{(0)}}{m}}\frac{k_z}{\Omega} \sim  {r_L
  k_\perp}}$, these equations simplify into
\begin{eqnarray}
&&\nabla_\perp \cdot \Pi_\perp
=   \frac{p_\perp^{(0)}}{2\Omega}{\widehat z}\times
\Delta_\perp u   - \frac{\rho^{(0)}}{2m}r_L^2
\Delta_\perp \nabla_\perp T_\perp^{(1)} - \frac{1}{2\Omega^2}
\Delta_\perp\nabla_\perp {\widetilde r_{\perp\perp}} +
\frac{1}{2\Omega}  {\widehat z}\times\partial_t \nabla_\perp \cdot \Pi_\perp    \label{Piperp}\\
&&S^\perp_\perp = \frac{2 p_\perp^{(0)}}{m\Omega} {\widehat z} \times\nabla_\perp T_\perp^{(1)} -
\frac{p_\perp^{(0)}}{2} r_L^2 \Delta_\perp u_\perp + \frac{2}{\Omega} {\widehat z} 
\times\nabla _\perp {\widetilde r_{\perp\perp}} 
+  \frac{1}{\Omega} {\widehat z}\times  \partial_t S^\perp  \label{Sperpperp}\\
&&\Pi_z =  \frac{\widehat z}{\Omega} \times \Big (\nabla_\perp S^\perp_z +
p_\perp^{(0)} \nabla_\perp u_z + p_\|^{(0)}
\partial_z u_\perp - ( p_\perp^{(0)} - p_\|^{(0)} ) \partial_t
{\widehat b_\perp} + \partial_t \Pi_z \Big ) \nonumber \\
&& \qquad - \frac{1}{\Omega^2} \partial_z \Big ( \frac{p_\perp^{(0)}}{m}\nabla_\perp T_\|^{(1)} - 
2 \frac{p_\perp^{(0)} -  p_\|^{(0)}}{m} T_\|^{(0)} \partial_z {\widehat
  b_\perp} + \nabla_\perp{\widetilde r_{\|\perp}} \Big )  \label{Piz}\\
&&S_\perp^\| = \frac{\widehat z}{\Omega} \times \Big (\frac{p_\perp^{(0)}}{m}\nabla_\perp T_\|^{(1)} - 
2 \frac{p_\perp^{(0)} -  p_\|^{(0)}}{m} T_\|^{(0)} \partial_z {\widehat
  b_\perp} + \nabla_\perp{\widetilde r_{\|\perp}}  +\partial_t  S_\perp^\|
\Big ) \nonumber \\
&& \qquad - \frac{2 T_\|^{(0)}}{m\Omega^2} \partial_z 
\Big ( \nabla_\perp S^\perp_z + p_\perp^{(0)} \nabla_\perp u_z + p_\|^{(0)}
\partial_z u_\perp - ( p_\perp^{(0)} - p_\|^{(0)} ) \partial_t
{\widehat b_\perp} \Big ) \label{Sparpperp}.
\end{eqnarray}

The last term in the r.h.s. of Eq. (\ref{Piperp}) can be consistently replaced
by $\displaystyle{ \frac{-1}{4\Omega^2} p_\perp^{(0)} \Delta_\perp \partial_t
  u_\perp}$, that in (\ref{Sperpperp}) by 
$\displaystyle{\frac{-2 p_\perp^{(0)}T_\perp^{(0)}}{m\Omega^2} \nabla_\perp
  \partial_t\Big ( \frac{T_\perp}{T_\perp^{(0)}}\Big )}$. A similar
substitution is made in Eqs. (\ref{Piz}) and (\ref{Sparpperp}), the terms
involving 
%a time derivative, $\nabla_\perp \cdot\Pi_\perp$, $S^\perp_\perp$, 
$\partial_t \Pi_z$ and $\partial_t S_\perp^\|$, being  replaced by their
leading order expressions within the  linear description.

\subsection{Simplified nonlinear equations for the longitudinal components of the heat flux
  vectors}
In deriving the dynamical equations governing the longitudinal
components of the heat flux 
vectors, we retain the coupling to the transverse components and to the 
gyroviscous tensor at the linear level only, because of the presence of a 
$1/\Omega$ factor, and the assumption that the present equations are restricted
to the description of the large scales. We retain the other
couplings that include quadratic contributions with respect to the
fluctuations (weak nonlinearity regime). Note that the
variation of ${\widehat b_z}$ has a magnitude 
that scales like the square of the perturbations. One then gets
\begin{eqnarray}
&&\partial_t  S_z^\| + \nabla\cdot (S_z^\| u) + 3 S_z^\|  \partial_z u_z 
+ 3 p_\| ({\widehat b}\cdot \nabla)\Big (\frac{p_\|}{\rho} \Big ) -
p_\perp {\widehat b_\perp}\cdot \nabla_\perp\Big (\frac{p_\|}{\rho}\Big)
\nonumber \\
&& +\frac{2p_\|}{\rho}(p_\|- p_\perp) \partial_z{\widehat  b_z}  
 + \nabla\cdot ({\widetilde r_{\|\|}} {\widehat b}) 
 - 3 {\widetilde r_{\|\perp}} \nabla\cdot {\widehat b} 
-(b_\perp\cdot \nabla_\perp) {\widetilde r_{\|\perp}}=0.
\end{eqnarray} 

Similarly, when considering the equation governing $\displaystyle{S_z^\perp}$,
one gets 
\begin{eqnarray}
&&\partial_t S_z^\perp + \nabla\cdot(u S_z^\perp) + S_z^\perp \nabla\cdot u +
p_\|({\widehat b}\cdot\nabla)\Big (\frac{p_\perp}{\rho}\Big ) - 2 p_\perp
({\widehat b_\perp}\cdot \nabla_\perp)\Big(\frac{p_\perp}{\rho} \Big )\nonumber \\
&&+\frac{p_\perp}{\rho}\Big (\partial_x \Pi_{xz} + \partial_y
  \Pi_{yz}\Big ) + \nabla \cdot ({\widetilde r_{\|\perp}} {\widehat b}) +
\Big ( \frac{p_\perp (p_\| -p_\perp)}{\rho} -{\widetilde r_{\perp\perp}}+
{\widetilde r_{\|\perp}}\Big ) (\nabla\cdot {\widehat b})\nonumber \\
&&-({\widehat b_\perp}\cdot \nabla_\perp){\widetilde r_{\perp\perp}}=0.
\end{eqnarray}

\section{Linear kinetic theory}

Let us assume that the equilibrium state is characterized for each particle species
by a bi-Maxwellian distribution function 
$\displaystyle{
f_{0} = \frac{1}{(2\pi)^{3/2}} \frac{m^{3/2}}{T^{(0)}_{\perp }
T^{(0)1/2}_{\| }}
\exp\Big \{-\Big (\frac{m}{2T^{(0)}_{\| }} v_\|^2 + 
\frac{m}{2T^{(0)}_{\perp }}  v_\perp^2\Big )\Big \}}$.
For small disturbances, the perturbation $f_1$ of the distribution function is linearly
expressed in terms of the parallel and transverse electric field components
that are conveniently written in terms of potentials, in the form
$ \displaystyle{E_z=-\partial_z \Psi}$ and 
$\displaystyle{E_\perp =-\nabla_\perp \Phi -\frac{1}{c}\partial_t A_\perp}$
with $\displaystyle{B = B_0 \hat z + \nabla \times A}$ 
and the gauge condition $\displaystyle{\nabla \cdot A = 0}$. We also denote by
$b_z$ the magnetic field fluctuations along the $z$-direction. 

The hydrodynamic moments are easily computed in a low frequency
expansion, retaining only contributions up to order
$\displaystyle{\frac{\omega}{\Omega}\sim
  \frac{k_z}{\Omega}\sqrt{\frac{2T_\perp^{(0)}}{m}}\ll 1}$, with no condition on 
$\displaystyle{\frac{k_\perp}{\Omega}\sqrt{\frac{2T_\perp^{(0)}}{m}}}$. 
Let us also introduce  $\displaystyle{b=\frac{T_\perp^{(0)} k_\perp^2}{m \Omega^2}= k_\perp^2
r_L^2}$, $\displaystyle{\zeta = \frac{\omega}{|k_z|}\sqrt{\frac{m}{2
    T_\|^{(0)}}}}$ and define  the functions 
$\Gamma_\nu(b) = e^{-b} I_\nu(b)$ in terms of the modified Bessel
function $I_\nu(b)$. A standard calculation leads to the following results in
terms of the plasma response function
$\displaystyle{R(\zeta)= 1 + \zeta Z(\zeta)}$, where 
$\displaystyle{Z(\zeta)}$ is the plasma dispersion function.

The longitudinal and transverse temperature  perturbations $T_\|^{(1)}$ and $T_\perp^{(1)}$
are given by
\begin{equation}
\frac{T_\|^{(1)}}{T_\|^{(0)}}= \Big (1-R(\zeta) + 2\zeta^2 R(\zeta)\Big ) 
\frac{T_\perp^{(0)}}{T_\|^{(0)}} \Big [ \Big (\Gamma_1(b) -\Gamma_0(b) \Big )
\frac{b_z}{B_0} - \Gamma_0(b) \frac{e\Psi}{T_\perp^{(0)}}\Big ]\nonumber\\
 \label{Tparal}
\end{equation}
and 
\begin{eqnarray}
&&\frac{T_\perp^{(1)}}{T_\perp^{(0)}}=\Big (\frac{T_\perp^{(0)}}{T_\|^{(0)}}
R(\zeta)-1 \Big )\Big (-2b\Gamma_1(b) + 2b\Gamma_0(b) -\Gamma_0(b)\Big )
\frac{b_z}{B_0}- \Big (b\Gamma_1(b)-b\Gamma_0(b) \Big)R(\zeta)
\frac{e\Psi}{T_\|^{(0)}}\nonumber \\
&& +  \Big (b\Gamma_1(b)-b\Gamma_0(b) \Big)
\frac{e}{T_\perp^{(0)}}\Big (\Phi + \frac{k_z^2}{k_\perp^2} (\Phi - \Psi)\Big). \label{Tperp}
\end{eqnarray}

When restricted to the linear approximation, the elements of the heat flux tensor reduce to 
${ q_{ijk}= n^{(0)} m \int v_i v_j v_k f_1 d^3v - 
u_i p^{(0)}_{jk} - u_j p^{(0)}_{ik} -  u_k p^{(0)}_{ij}}$.
For  the flux vectors
${S^\|_i = q_{ijk} {\widehat b_j}{\widehat b_k}}$
and ${S^\perp_i = \frac{1}{2} q_{ijk} (\delta_{jk}-{\widehat
    b_j}{\widehat b_k})}$, 
one then has  ${S_i^\|=  n^{(0)} m\int v_i v_\|^2 f_1 d^3v - 
p_\|^{(0)}( u_i + 2 \delta_{i3} u_z)}$ and
${S_i^\perp = \frac{ n^{(0)} m}{2} \int v_i v_\perp^2 f_1 d^3v - 
 p_\perp^{(0)}( 2u_i -  \delta_{i3} u_z)}$. 

It results that
\begin{equation}
S_z^\| = - p_\|^{(0)}  \frac{T_\perp^{(0)}}{T_\|^{(0)}} 
\frac{\omega}{k_z} \Big (1 -3 R(\zeta) + 2 \zeta^2 R(\zeta)\Big ) 
\Big [\Big (\Gamma_0(b) - \Gamma_1(b) \Big )  \frac{b_z}{B_0}
 + \Gamma_0(b)\frac{e\Psi}{T_\perp^{(0)}}\Big ]
\end{equation}
and
\begin{eqnarray}
&&S_z^\perp =  p_\perp^{(0)} \Big \{ \frac{T_\perp^{(0)}}{T_\|^{(0)}}
  \frac{\omega}{k_z} \Big ( 2 b \Gamma_0(b) - \Gamma_0(b) -2b \Gamma_1(b)\Big ) R(\zeta)
\frac{b_z}{B_0} 
+ \frac{\omega}{k_z} b \Big (\Gamma_0(b) - \Gamma_1(b) \Big ) R(\zeta)
\frac{e\Psi }{T_\|^{(0)}} \nonumber \\
&&  - \frac{T_\perp^{(0)} - T_\|^{(0)}}{m}
  \frac{k_z}{\omega}  \Big (  b \Gamma_0(b) - b  \Gamma_1(b)
  \Big ) \frac{e}{T_\perp^{(0)}} \Big ( 1 +  \frac{k_z^2}{k_\perp^2} \Big ) (\Phi -
  \Psi) \Big \}.
\end{eqnarray}

For a gyrotropic equilibrium distribution function, symmetric in the direction
of the ambient field, the elements of the fourth  order moment perturbation
read  ${ r_{ijkl}^{(1)}= n^{(0)} m \int v_i v_j v_k v_l f_1 d^3v}$.
One computes the scalar quantities  $ {r_{\|\|}^{(1)}= r_{ijlk}^{(1)} {\widehat b_i}{\widehat
  b_j}{\widehat b_k}{\widehat b_l}=  n^{(0)} m \int v_\|^4 f_1 d^3v}$, 
 $r_{\|\perp}^{(1)}=  {\frac{1}{2} r_{ijlk}^{(1)}
(\delta_{ij} - {\widehat b_i}{\widehat  b_j}){\widehat b_k}{\widehat b_l} 
=\frac{1}{2}n^{(0)} m \int v_\|^2 v_\perp^2 f_1 d^3v}$
and  $r_{\perp\perp}^{(1)}=   {\frac{1}{4}r_{ijlk}^{(1)}
(\delta_{ij} - {\widehat b_i}{\widehat  b_j})(\delta_{lk} - {\widehat
  b_k}{\widehat  b_l}) = \frac{1}{4} n^{(0)} m \int v_\perp^4 f_1 d^3v}$.
 After linearization of Eqs. (\ref{rparpar})--(\ref{rperpperp}) one gets
\begin{eqnarray}
&&{\widetilde r_{\|\|}} =  \frac{p_\|^{(0)} T_\perp^{(0)}}{m} \Big [
2 \zeta^2\Big (1 + 2 \zeta^2 R(\zeta)\Big ) + 3 \Big (R(\zeta) -1\Big ) - 12 \zeta^2 R(\zeta)\Big
] \Big [\Big (\Gamma_1(b)-\Gamma_0(b)\Big ) \frac{b_z}{B_0} -
\Gamma_0(b)\frac{e\Psi}{T_\perp^{(0)}} \Big ]  \nonumber \\ 
\\  
&&{\widetilde r_{\|\perp}} = \frac{{p_\perp^{(0)}}^2}{\rho^{(0)}} \Big ( 1 -
  R(\zeta) + 2 \zeta^2 R(\zeta) \Big ) \Big [\Big (2 b \Gamma_0(b) - \Gamma_0(b) - 2
  b \Gamma_1(b)\Big ) \frac{b_z}{B_0} + b \Big (\Gamma_0(b) -
  \Gamma_1(b)\Big )\frac{e\Psi}{T_\perp^{(0)}}\Big ]\\
&&{\widetilde r_{\perp\perp}} = \frac{{p_\perp^{(0)}}^2}{\rho^{(0)}} 
\Big \{ 
\Big (4b^4\Gamma_1(b) - 4b^2\Gamma_0(b) -b\Gamma_1(b) +3b \Gamma_0(b)\Big )
\Big (\frac{T_\perp^{(0)}}{T_\|^{(0)}}R(\zeta)  -1 \Big ) \frac{b_z}{B_0}\nonumber \\
&& \qquad + \Big (2b^2\Gamma_1(b) + b\Gamma_1(b) -
2b^2\Gamma_0(b)\Big ) R(\zeta)\frac{e\Psi}{T_\|^{(0)}} + \Big (2b^2\Gamma_0(b)
-6 b
\Gamma_1(b)\Big ) \frac{e}{T_\perp^{(0)}} \Big ( \Phi +
\frac{k_z^2}{k_\perp^2}(\Phi-\Psi)\Big ) \Big \}. \nonumber \\
\end{eqnarray}

\section{A Landau fluid closure}

When  comparing the expression of ${\widetilde r_{\|\|}^{(1)}}$ with those of 
$S_z^{\|}$ or $T_\|^{(1)}$ provided by the kinetic theory, one gets 
\begin{equation}
{\widetilde r_{\|\|}} = \sqrt{\frac{2T_\|^{(0)}}{m}}
\frac{2 \zeta^2\Big (1 + 2 \zeta^2 R(\zeta)\Big ) + 3 \Big (R(\zeta) -1\Big ) - 12 \zeta^2
  R(\zeta)}{2\zeta {\rm sgn \,} (k_z) \Big (1 - 3 R(\zeta) + 2 \zeta^2
  R(\zeta)\Big )} S_z^{\|} \equiv \sqrt{\frac{2T_\|^{(0)}}{m}} {\cal
  F}_S S_z^{\|}.
\end{equation}
and 
\begin{equation}
{\widetilde r_{\|\|}} =  \frac{p_\|^{(0)}T_\|^{(0)}}{m} 
\ \frac{2 \zeta^2\Big (1 + 2 \zeta^2 R(\zeta)\Big ) + 3 \Big (R(\zeta) -1\Big ) - 12 \zeta^2
  R(\zeta)}{1-R(\zeta) + 2\zeta^2 R(\zeta)} \ \frac{T_\|^{(1)}}{T_\|^{(0)}}
\equiv\frac{p_\|^{(0)}T_\|^{(0)}}{m}  {\cal F}_T \frac{T_\|^{(1)}}{T_\|^{(0)}}. 
\end{equation}
One then notices that when replacing the plasma response function $R$ by its four pole
approximant 
$$\displaystyle{R_4(\zeta) ={\frac {4-2\,i\sqrt {\pi
      }\zeta+ \left( 8-3\,\pi  \right) {\zeta}^{2}}{4-6\,i
\sqrt {\pi } \zeta+ \left( 16-9\,\pi  \right)
      {\zeta}^{2}+4\,i\sqrt {\pi } {\zeta}^{
3}+ \left( 6\,\pi -16 \right) {\zeta}^{4}}}},
 $$
one has the identity 
\begin{equation}
\lambda \frac{{\cal F}_S}{{\cal F}_T} + i \mu \frac{k_z}{|k_z|}= {\cal F}_S
\end{equation}
with  $\displaystyle{\lambda = \frac{32- 9\pi}{3\pi -8}}$
and $\displaystyle{\mu= \frac{-2\sqrt{\pi}}{3\pi -8}}$. This leads to the
closure relation 
\begin{equation}
{\widetilde r_{\|\|}}= \lambda p_\|^{(0)}\frac{T_\|^{(0)}}{m}
\frac{T_\|^{(1)}}{T_\|^{(0)}} +  \mu \sqrt{\frac{2T_\|^{(0)}}{m}} \frac{i k_z}{|k_z|}S_z^\|,
\end{equation}
which identifies with Eq. (34) of  Ref. [8]. Note that this
closure is here established with no assumption on the magnitude of
the transverse wavenumbers.

On the other hand, ${\widetilde r_{\|\perp}}$
 can be expressed in terms of $S_z^{\perp}$ and the parallel current
$\displaystyle{j_z}$. One has
\begin{eqnarray}
&&{\widetilde r_{\|\perp}} = \sqrt{\frac{2T_\|^{(0)}}{m}} \frac{1 - R(\zeta) + 2
  \zeta^2 R(\zeta)}{2\zeta R(\zeta)} \Big [S_z^\perp + \Big (
   \Gamma_0(b) - \Gamma_1 (b)\Big ) \frac{p_\perp^{(0)} p_\|^{(0)}}{\rho^{(0)}v_A^2}\Big
  (\frac{T_\perp^{(0)}}{T_\|^{(0)}} -1 \Big )  \frac{j_z}{en^{(0)}}
   \Big ]
\end{eqnarray}
where $v_A = B_0/\sqrt{4\pi \rho^{(0)}}$ is the Alfv\'en velocity and
$\rho^{(0)}$ the plasma density at equilibrium.

When dealing with ${\widetilde r_{\|\perp}}$, the approximation
consisting in replacing the plasma response function $R$ by its two pole
Pad\'e approximant $\displaystyle{R_2(\zeta)= 1/(1-i \sqrt {\pi}\zeta 
  - 2 \zeta^2)}$, as performed to obtain Eq. (35) of Ref. [8] is
not satisfactory since it does not correctly reproduce the large $\zeta$
decay of the imaginary part of the fraction $\displaystyle{
  \frac{1 - R(\zeta) + 2  \zeta^2 R(\zeta)}{2\zeta R(\zeta)}}$. 
Similar possible overestimate of the Landau damping by Landau fluid
models are mentioned in Ref. [16]. In contrast,
using $\displaystyle{R_3(\zeta) =
\frac{2 - i \sqrt{\pi} \zeta} {2 - 3 i \sqrt{\pi} \zeta - 4
\zeta^2 + 2 i \sqrt{\pi}\zeta^3}}$, one has the approximation
$\displaystyle{ \frac{1 - R(\zeta) + 2
  \zeta^2 R(\zeta)}{2\zeta R(\zeta)} \approx \frac{i \sqrt{\pi}} {-2 + i
  \sqrt{\pi} \zeta}}$. This leads to write the evolution equation
\begin{equation}
\Big (\frac{d}{dt} - \frac{2}{\sqrt{\pi}}\sqrt{\frac{2T_\|^{(0)}}{m} }{\cal
  H}_z\partial_z \Big )
{\widetilde r_{\|\perp}} + \frac{2T_\|^{(0)}}{m} \partial_z
\Big [S_z^\perp + 
\frac{p_\perp^{(0)} }{v_A^2} \Big (\frac{T_\perp^{(0)}- T_\|^{(0)}}{m_p}   \Big )
\frac{ j_z}{e n^{(0)} } \Big ]=0,
\end{equation}
where in the large-scale limit we are here concerned with, we made the expansion
$\displaystyle{b\Gamma_0(b) - b\Gamma_1(b) \approx b =
  k_\perp^2 r_L^2}$. 
The notation $m_p$ is used  in situations where the proton mass is to 
remain unchanged when turning to the corresponding equation for the electron.
In Fourier space, the Hilbert transform ${\cal H}_z$ reduces to the
multiplication by $i \,{\rm  sgn}\, k_z$.
The convective derivative has been reintroduced to ensure
Galilean invariance.

Finally, the reduced moment ${\widetilde r_{\perp\perp}}$ turns out to be
totally negligible at large scales and will thus not be retained.

\section{Comments on the resulting model}

The equations derived above for the ions are  easily adapted to the
electrons for which they  greatly simplify when making the approximation
$m_e/m_p \ll 1$. This leads to neglect the non gyrotropic components of the 
corresponding pressure tensor. Note that the transverse components of the electron
heat flux
vectors survive due to the contributions of terms involving the product $m_e
\Omega_e$ (see Section II.F).
The system is to be supplemented by Faraday equation 
and Amp\`ere's law where the displacement current is neglected.  
In this two-fluid formulation, energy is conserved, as discussed by
Ramos. \cite{R05} It might nevertheless be advantageous to filter out the scales
associated with electrostatic waves by prescribing quasi-neutrality, 
replacing the electron momentum equation by a generalized Ohm's law, 
and turning to a one-fluid description. Numerical simulations of a monofluid 
model obtained from a simplified 
version of the present model have shown that energy 
is in practice very well conserved. \cite{BPS04}

When compared with the previous model \cite{PS04} designed to reproduce the oblique
Alfv\'en wave dynamics, the present approach proves to be more systematic and,
as discussed  below, allows one to accurately simulate all dispersive 
MHD waves, including oblique and
transverse magnetosonic waves (see Section VI). The previous model has on the other hand 
the advantage of including a nonlinear description of the gyroviscous tensor. 
It is of interest to see how, when linearized and restricted to the case of the 
Alfv\'en wave scaling (also neglecting the gyroviscous tensor contribution), 
the equations governing the gyrotropic heat
fluxes in the present model compare with those of the previous one. It turns
out that Ref. [12] unfortunately includes a few
algebraic errors originating from a sign error leading to an incorrect factor
3 in Eq. (C.8),  a missing multiplicative factor $m_p/m_r$ in the r.h.s. of
Eqs. (C.9) and (C.10) and  a missing minus sign in front of the first occurrence
of $\Omega_p/\Omega_r$ in Eq. (C.12). This in particular affects the equations for the
gyrotropic heat fluxes where the contribution $v_{\Delta e}^2$ in the r.h.s. of 
Eq. (56) should be suppressed, the square bracket in Eq. (66)
replaced by $[v_{\Delta r}^2 {\rm sgn} \, q_r - v_A^2 (\delta_{rp}-1) -v_{th,r}^2
\delta_{rp}]/v_A^2$
and the factor 3 in the last term in the r.h.s. of Eq. (67) also discarded.
 After correcting these errors and taking into account that pressure and heat flux tensors were
computed using barycentric velocities, one easily checks that the parallel 
heat flux equation is exactly recovered and that the equations for the
perpendicular heat flux of both models identify in the isothermal limit where
the time derivatives are negligible. This limitation  originates
from the insufficient order of the 
Pad\'e approximant used in the previous model.

\section{MHD wave dynamics}
 
When restricted to a one or quasi one-dimensional dynamics along the ambient
field, only the longitudinal components of the parallel and transverse heat
flux vectors (that correspond to the gyrotropic contributions to the heat flux
tensor) arise in the equations of motion.  
A  long-wave reductive perturbative expansion performed on the resulting
Landau-fluid model reproduces the kinetic derivative nonlinear Schr\"odinger
equation derived from the VM equations for Alfv\'en waves with a
typical length scale large compared with the ion Larmor radius, \cite{PS03a} 
up to the replacement of the plasma response function by  appropriate
Pad\'e approximants. As a consequence, the modulational type
instabilities (including filamentation\cite{PS03c}) of Alfv\'en waves and their weakly
nonlinear developments are correctly reproduced. \cite{PS03b} Numerical
simulations of such regimes are presented in Ref. [10] where a
study of the decay instability is also presented and
validated by comparison with hybrid simulations. \cite{Va95}

As stressed in  Ref. [13], the correct determination of the
dispersion relation for transversally 
propagating magnetosonic waves requires a detailed description
of non-gyrotropic contributions to the pressure and heat flux
tensors. When restricted to a purely transverse dynamics, the
present model reduces to the fluid model used in Ref. [13]  that
exactly reproduces the large-scale kinetic theory (note that a factor
$3/2$ is missing in front of the $z$-term in $\epsilon_{xy}$ given in
Eq. (2.8) of the latter reference).

The present model easily reproduces the
dispersion relation for kinetic Alfv\'en waves (KAW) for which the
crucial ingredient is the contribution to the transverse velocity
originating from the time derivative of the leading  order
gyroviscous stress [last term in
  Eq. (\ref{Piperp})]. \cite{MDH96,PS03d,PS04} Whereas  
these KAWs are also captured by a low frequency expansion of
the kinetic equations, \cite{HaCh76,CJ99} this is not the case for oblique Alfv\'en
waves. The reason is that an expansion at order $\omega/\Omega$
includes contributions of order $k_\perp^2 r_L^2$ when $k_z/k_\perp$
scales like $k_\perp r_L$ as for KAWs, but only includes terms of
order $k_\perp r_L$ for finite angles of propagation. The same
limitation holds for the gyrokinetic formalism. The present fluid
formalism however allows one to obtain the correct linear dynamics for
oblique Alfv\'en waves, as was shown in Ref. [21], using a
Landau fluid model actually contained in the present one.

\section{Concluding remarks}

We have constructed a Landau fluid model that reproduces all 
large-scale dispersive MHD waves in a warm collisionless plasma. This model
may be most useful not only  for 
numerical simulations involving a broad range of scales, but also
for analytic purposes, such as the computation of secondary
instabilities. An example is provided by the filamentation
instability of parallel propagating Alfv\'en waves. This mechanism
may be relevant in the understanding of the evolution of Alfv\'en
waves in magnetospheric plasmas that often display very filamentary
structures. \cite{AMML04} The present model allows one to account for linear Landau
damping, dominant FLR corrections as well as drift velocities, that play an
important role in these plasmas whose equilibrium state often involves  a 
large scale longitudinal current.  
The importance of nonlinear kinetic effects such as particle trapping
that are here neglected should be estimated by comparison with fully
kinetic simulations.

In a sufficiently anisotropic plasma, the mirror instability can
develop, whose threshold is 
accurately reproduced by the present fluid model. \cite{SHD97,BPS04}
A difficulty nevertheless 
originates in that, for large-scale mirror modes,
the growth rate of perturbations propagating in the most unstable direction
scales like the transverse wave number of
the perturbation, which makes the smallest scales retained in a large-scale simulation
to be the most unstable. The instability actually reaches a maximal rate at a
scale comparable to the ion Larmor radius and is arrested at smaller scales,
under the effect of FLR corrections. \cite{PSBT04}  Small transverse scales are thus to be retained.
A promising approach consists in expressing, at the level of the linear kinetic
theory, non-gyrotropic contributions in a closed form suitable 
for being incorporated into fluid equations. Explicit  reference to the
plasma response function should in particular be eliminated.
A model  that reproduces the arrest of the
mirror instability and that is  simple enough to permit accurate
numerical simulations will be presented in a forthcoming paper.\cite{PS05}

\begin{acknowledgments}
This work benefited of support from CNRS  programs 
``Soleil-Terre'' and ``Physique et Chimie du Milieu Interstellaire''.
\end{acknowledgments}

\end{document}